\documentclass[fleqn,usenatbib]{mnras}

\usepackage{newtxtext,newtxmath}
\usepackage{fix-cm}
\usepackage[T1]{fontenc}

\DeclareRobustCommand{\VAN}[3]{#2}
\let\VANthebibliography\thebibliography
\def\thebibliography{\DeclareRobustCommand{\VAN}[3]{##3}\VANthebibliography}

\usepackage{float}
\usepackage{graphicx}	%
\usepackage{amsmath}	%
\usepackage{dirtytalk}
\usepackage{siunitx}
\usepackage[dvipsnames]{xcolor}

\DeclareSIUnit{\hred}{\textit{h}}
\DeclareSIUnit{\Msun}{M_\odot}

\usepackage{cleveref}
\crefname{equation}{Eq.}{Eqs.}
\crefname{figure}{Fig.}{Figs.}
\crefname{table}{Table}{Tables}
\crefname{section}{Section}{Sections}
\crefname{appendix}{Appendix}{Appendices}
\Crefname{equation}{Eq.}{Eqs.}
\Crefname{figure}{Fig.}{Figs.}
\Crefname{table}{Table}{Tables}
\Crefname{section}{Section}{Sections}
\Crefname{appendix}{Appendix}{Appendices}

\usepackage[normalem]{ulem}

\newcommand{\report}[1]{\begin{color}{black}#1\end{color}}

\newcommand{\Jfil}{{\ensuremath{\mathbf{J}, \mathbf{e}_3}}}
\newcommand{\Afil}{{\ensuremath{\mathbf{A}, \mathbf{e}_3}}}
\title[Nature vs. Nurture of DM Halos]{\report{Exploring} the causal effect of cosmic filaments on dark matter halos}

\author[A. Storck et al.]{
Anatole Storck,$^{1,2}$\thanks{E-mail: storckanatole@gmail.com}
Corentin Cadiou,$^{1,3}$
Oscar Agertz,$^{1}$ and
Daniela Galárraga-Espinosa$^{4,5}$
\\
$^{1}$Lund Observatory, Division of Astrophysics, Department of Physics, Box 43, SE-221 00, Lund, Sweden\\
$^{2}$Sub-department of Astrophysics, University of Oxford, DWB, Keble Road, Oxford OX1 3RH, United Kingdom\\
$^{3}$Institut d’Astrophysique de Paris, UMR 7095, CNRS, and Sorbonne Université, 98 bis boulevard Arago, 75014 Paris, France\\
$^{4}$Max-Planck Institute for Astrophysics, Karl-Schwarzschild-Str. 1, 85741 Garching, Germany\\
$^{5}$Kavli IPMU (WPI), UTIAS, The University of Tokyo, Kashiwa, Chiba 277-8583, Japan}

\date{Accepted XXX. Received YYY; in original form ZZZ}

\pubyear{2025}

\begin{document}
\label{firstpage}
\pagerange{\pageref{firstpage}--\pageref{lastpage}}
\maketitle

\begin{abstract}
The way in which the large-scale cosmic environment affects galactic properties is not yet understood. Dark matter halos, which embed galaxies, initially evolve following linear theory. Their subsequent evolution is driven by non-linear structure formation in the halo region and in its outer environment.
In this work, we present the first study where we explicitly control the linear part of the evolution of the halo, thus revealing the role of non-linear effects on halo formation.
We focus specifically on the effect of proximity to a large cosmological filament.
We employ the splicing method to keep fixed the initial density, velocity, and potential fields where a halo will form while changing its outer environment, from an isolated state to one where the halo is near a large filament.
In the regime of Milky Way-mass halos, we find that mass and virial radius of such halos are not affected by even drastic changes of environment, whereas
halo spin and shape orientation with respect to a massive filament is largely impacted, with fluctuations of up to $\SI{80}{\percent}$ around the mean value. Our results suggest that halo orientation and shape cannot be predicted accurately from a local analysis in the initial conditions alone.
This has direct consequences on the modeling of intrinsic alignment for cosmic shear surveys, like Euclid.
Our results highlight that non-linear couplings to the large-scale environment may have an amplitude comparable to linear effects, and should thus be treated explicitly in analytical models of dark matter halo formation.
\end{abstract}

\section{Introduction}\label{seq:Introduction}

In the standard picture of structure formation, galaxies form in the centers of dark matter (DM) halos \citep[e.g.][]{White1978, Blumenthal1984, White1991}. The physics setting the properties of DM halos is a central topic of galaxy formation theory, with seminal work carried out to understand the origins of their internal structure \citep[e.g.][]{Navarro1991, Navarro1996, Moore1999, Bullock2001}, role of background cosmology \citep[e.g.][]{maccio08} and cosmic environment \citep[e.g.][]{lemson99,Hahn2007a,Cautun2014,musso2018,Hellwing2021}.
With the advent of Stage IV-surveys such as Euclid, the need to understand how galaxy and halo properties emerge has never been more pressing.

Many models and theories have been brought forward to predict current day halo properties from the early-Universe initial conditions,
where density fluctuations are small enough to be treated with linear perturbation theory (`the linear regime').
This allows to relate cosmological models to (indirectly) observable properties.
Analysis of these perturbations allow predictions of the statistics of the collapse of DM halos using extended Press-Schechter formalisms \citep{PressSchechter,Bond1991}. This approach is complemented by the peak-patch approach \citep{BBKS1986,Bond1996} that associates peaks in the initial conditions to halos in the evolved Universe, while merger rates can be obtained by counting the coalescence of these peaks \citep[the critical-event theory,][]{Hanami2001,Cadiou2020}.
The statistical properties of mergers can then be predicted analytically through extensions of the aforementioned theories \citep{Lacey1993,Neistein2008,Cadiou2023} or, alternatively, it can be measured in peak-patch simulations \citep{Stein2019}.
Conversely, the build up of halo angular momentum up to turnaround can be modeled analytically using tidal torque theory \citep[TTT,][]{PeeblesSpin, Doroshkevich1970, White1984, Catelan1996, Crittenden2001, Schafer2009,Lopez2024}.

\begin{figure*}
	\includegraphics[width=\textwidth]{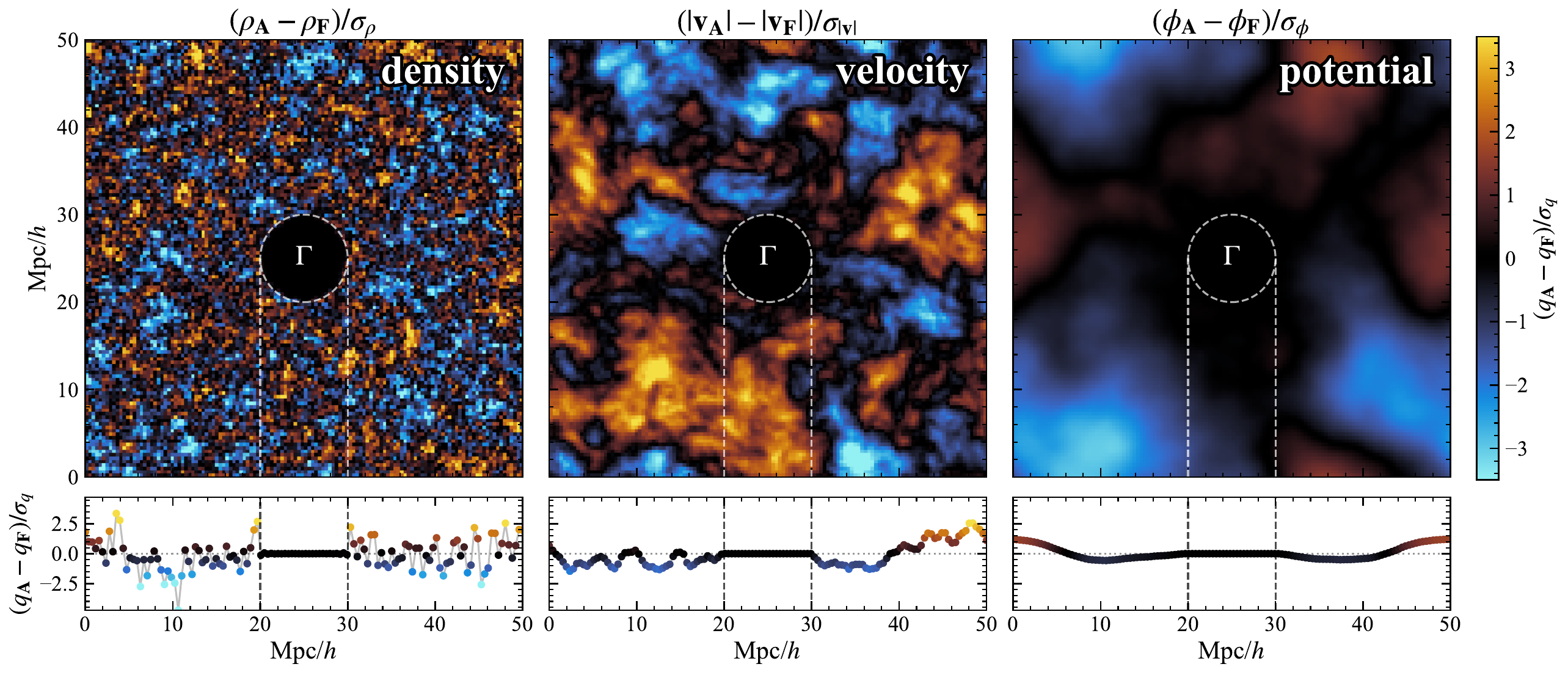}
    \caption{Difference ($\boldsymbol{A} - \boldsymbol{F}$) between two sets of cosmological initial conditions where one set $\boldsymbol{A}$ has been randomly generated, and the other $\boldsymbol{F}$ has been spliced from the former into a different random field with a sphere of radius $r=\SI{5}{Mpc/\hred}$ positioned in the center of the box, denoted $\boldsymbol{\Gamma}$. The top row shows the center slice of the box in density, velocity, and potential fields, where the colormap has been normalized by dividing out the standard deviation of the combined field. We take the magnitude of the velocity vectors as our scalar velocity field for visualization. The bottom row follows the horizontal line in the slice which intersects the center of the splicing region. The two dashed lines in each plot represent the spliced region. The black color indicates that the two fields are the same, which we see occurs at all points within the spliced region in all physical fields. For density splicing, the velocity and by extension potential are changed in the spliced region, so only the left column holds. See \cref{appendixg:contamination} for the difference ($\boldsymbol{B} - \boldsymbol{F}$).
    Our extension of splicing thus allows us to arbitrarily fix the density, velocity and potential fields within an arbitrary region in $\Lambda$CDM-compatible initial conditions.
    }
    \label{fig:splicing_fig}
\end{figure*}
Despite their success at predicting population statistics, these analytical approaches remain limited in their ability to accurately predict the properties of individual halos \citep[see e.g.][for spin prediction]{porciani_TestingTidaltorqueTheory_2002a}, casting doubt in the fundamental feasibility of such enterprise.
In recent years, progress has however been made to predict accurately mass profiles \citep{Lucie-Smith2019,lucie-smith_InsightsOriginHalo_2022,brownUniversalModelDensity2022} from the initial conditions, revealing that different components of the profile respond to different time- and spatial-scales in the initial conditions. Conversely, more complex late-time properties were found to respond to perturbations of the initial conditions (halo concentration, \citealt{splicing}; angular momentum, \citealt{cadiouAngularMomentumEvolution2021,cadiouStellarAngularMomentum2022}; or even stellar mass and kinematics, \citealt{moonMutualInformationGalaxy2024}).
However, the physical origin of these correlations and the amplitude of the responses remain to be pinned down.

In particular, one interesting property of DM halos known to be directly affected by the large scale structure (LSS) is the orientation of the halo's angular momentum relative to the nearest cosmological filament's direction
\citep[][and references therein]{Hahn2007b, CosmicBalletI, CosmicBalletII, CosmicBalletIII}. A similar effect known as intrinsic alignment affects galaxies, with such effect acting as a contaminant in weak lensing surveys \citep{Hirata2004}.
Numerical studies of the mass dependence on the spin orientation of DM halos to filament spine \citep[e.g.][]{AragonCalvo2007, Hahn2007a, Codis2012, Laigle2015, CosmicBalletI, CosmicBalletIII} show that halos at $z=0$ below (above) the spin-flip mass of $\sim 10^{11}-10^{12}{\rm M_\odot}$ tend to be preferentially aligned (perpendicular) to the filament.
\report{This effect is well explained qualitatively by conditional tidal torque theory \citep{Codis2015}.}
The spin-flip mass shows a strong redshift dependence \citep{Codis2018, CosmicBalletIII}, moving to lower masses for higher redshifts.
However, such numerical simulations are inherently uncontrolled on the scale of individual halos due to the diversity of Gaussian random fields (GRFs) from which they stem, preventing us from disentangling halo-to-halo variations from environmental effects.

\report{This paper aims at overcoming the limitations of diverse GRFs: we design a controlled numerical experiment where we explicitly fix the initial density, velocity and potential field of a halo while systematically changing its large-scale environment. We aim to understand how changing the fields outside of the Lagrangian patch of a given DM halo impacts its evolution. This is done using the splicing method. Fixing the initial density, velocity and potential fields, effectively sets all predictions from local linear theory (notably extended Press-Schechter and (conditional) TTT), and thus allows us to uncover which DM halo properties are \emph{non-linearly} coupled to their large-scale environment, and to quantify the amplitude of such coupling.
}

\report{Given the numerically expensive nature of this experiment, we focus on a small number of systems in the mass range $1.7-\SI{3.5e12}{\Msun}$ consistent with an expected misalignment signal to close filaments, enough to have a first sense of what is the direct variation of halo properties induced by environmental changes. We warn the reader that we do not aim at reproducing the population trends of \SI{e12}{\Msun} halos. Rather, our goal is to shed light on the driver of halo-by-halo fluctuations that is otherwise marginalized over in statistical studies. We do so by quantifying the changes in halo properties induced by a changing environment on a few handpicked halos.
}

This paper is organized as follows: in \cref{seq:Method}, we present the technique employed to build our experiment --~the splicing technique extended to the potential field~-- and how we apply it to DM halos. We describe our analysis of the resulting simulations and introduce the calculated halo properties in \cref{seq:Analysis}, and share such results in \cref{seq:Results}. Finally, we discuss the consequences of our findings in terms of the trends as a function of distance, how they differ from the linear models, and the prospects to future applications of splicing in \cref{seq:Discussion&Conclusions}.

\section{Method}\label{seq:Method}

We present five suites of genetically modified Milky Way-mass dark matter halos, targeting changes in the initial conditions through splicing \citep{splicing} such that the distance of the halo at $z=0$ to its closest major filament is systematically shifted further and further away. A description of the splicing technique is given in \cref{subseq:Splicing the potential field}, we detail our simulation setup in \cref{subseq:Simulations}, and detail our process to generate our initial conditions, including our selection criteria for the the halos and the filament  in \cref{subseq:Halo and filament selection}.

\subsection{Potential field splicing}\label{subseq:Splicing the potential field}

Splicing is a technique which takes a physical region $\Gamma$ in the ICs (positions and velocities of dark matter particles at $z\approx1100$) of one realization (the environment) and replaces it with the ICs of another realization (the halo).
The original splicing implementation operates on the initial density field \citep{splicing}, thereby setting the positions of all particles in the selected region exactly. Fixing the density alone however allows the initial velocities (or potential) to vary\footnote{Indeed, it only enforces $\delta = \partial_i v_i$ to be a constant, allowing the different components of the velocity to change.}, resulting in slightly different initial tides in the spliced region.
In this work, we extend the original implementation and splice the potential field (up to a constant) of the halo's Lagrangian patch. This not only sets the density field, but the velocity field as well. As a consequence, splicing the potential field fixes all \emph{local} information commonly used to predict DM halo properties (velocity, tides, density and its derivatives). Fixing the potential fully sets the linear part of the evolution of the halo and allows us to comparatively reveal the effect of non-linear coupling of the forming dark matter halo with collapsing large-scale structures outside the halo's patch. In other words, we are able to disentangle the halo's evolution due to the intrinsic properties of its Lagrangian patch from the evolution driven by non-linear coupling to the halo's environment.

Here, we describe the splicing method following the steps laid in \citet{splicing} with our added extension to the potential, $\phi$. The potential field  in Fourier space, $\tilde\phi_k$, is built up from a seeded Gaussian white noise field $\tilde{\mathbf{n}}_k$ and the potential covariance matrix $\textbf{C}_\phi$ derived from the dark matter density power spectrum $P(k)$ with each $k$ mode for the field being\footnote{Since the potential is defined up to a constant, we set $P(k=0)=0$. This also prevents division-by-zero errors in \cref{eq:phi_definition}.}
\begin{equation}
    \tilde\phi_k = \textbf{C}_\phi^{0.5} \tilde{\textbf{n}}_k, \quad \text{with} \quad \textbf{C}_\phi(k_i, k_j) = \delta_{\rm D}(k_i, k_j) \frac{P(k_i)}{k^4_i},
    \label{eq:phi_definition}
\end{equation}
where $\delta_{\rm D}$ is the Kronecker delta.
For our two sets of ICs, we have $\phi_{\rm env}$ and $\phi_{\rm halo}$ made up from their respective white noise fields.
We introduce the mask matrix $\textbf M$ which zeroes all points of a field lying outside of $\Gamma$, as well as its conjugate matrix $\overline{\textbf M}$ zeroing all points inside of $\Gamma$.
We build up the spliced potential field $\phi_{\rm spl}$ in real space as follows
\begin{equation}
    \phi_{\rm spl} = \overline{\textbf M}\phi_{\rm \alpha} + \textbf{M} (\phi_{\rm halo} - \phi_{\rm env}) + \phi_{\rm env}.
    \label{eq:spliced_field}
\end{equation}
The equation ensures that the spliced field, $\phi_{\rm spl}$, is equal to the halo field,  $\phi_{\rm halo}$, in $\Gamma$. Outside of it, $\phi_{\rm spl}$ recombines continuously to the environment,  $\phi_{\rm env}$, as quickly as the $\Lambda\rm CDM$ power spectrum allows.
\report{In order to set the ICs of the halo Lagrangian patch exactly, the outer potential (and density) fields need to change in some non-trivial way through $\phi_{\rm \alpha}$, due to the correlated nature of $\Lambda$CDM fields.}
$\phi_{\rm \alpha}$ is found by solving
\begin{equation}
    \textbf{C}_\phi^{0.5} \overline{\textbf M} \textbf{C}_\phi^{-1} \overline{\textbf M} \phi_{\rm \alpha} - \textbf{C}_\phi^{0.5} \overline{\textbf M} \textbf{C}_\phi^{-1} \textbf{M} (\phi_{\rm env} - \phi_{\rm halo}) = 0.
    \label{eq:minimization}
\end{equation}
Once the potential is computed, we obtain the velocity field and density field \emph{via} finite differences.
While solving \cref{eq:minimization} outright is impractical as it would involve inverting a $N^3\times N^3$ matrix\footnote{Although $\textbf{C}_\phi$ is diagonal in Fourier space and $\overline{\textbf M}$ in real space, their product isn't a sparse matrix.}, with $N$ our grid resolution, an approximate solution to the linear problem can be found iteratively \citep[in our case, with the {\sc MINRES} method\footnote{This is a suitable method, as the problem is symmetric when applied on the white noise. We also found {\sc MINRES} to converge in fewer iterations than the more classical conjugate-gradient method.}, see][]{MINRESpaige1975}.
We stop the minimization when the residuals become no larger than $10^{-7}$ the potential field. This proves sufficient to ensure smoothness of the solution at the boundary of the spliced region for the potential and its derived fields.

\begin{figure*}
	\includegraphics[width=\textwidth]{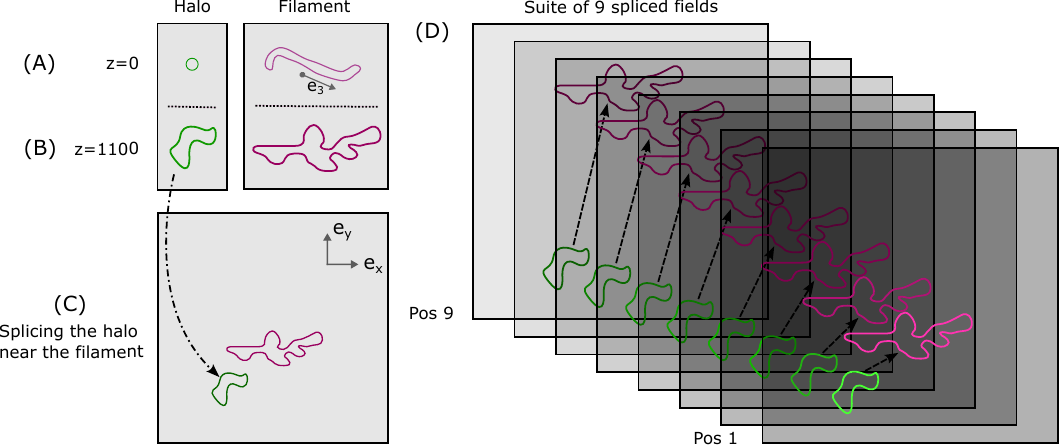}
    \caption{
    Schematic of the splicing operation between filament and halo, where each box is a slice in $e_z$ at the location of the proto-halo and proto-filament patches. \textbf{(A)} We first identify the filament and halo at present day in two respective simulations and \textbf{(B)} trace them back to their respective Lagrangian patches. \textbf{(C)} We then combine the halo's Lagrangian patch with the filament's using splicing (see the text for details). \textbf{(D)} We repeat this last operation 9 times for each halo, varying the distance between proto-halo and proto-filamentary patches, as illustrated by black dotted arrows.
    Since splicing preserves exactly the gravitational acceleration, tides and density in the halo's Lagrangian patch, our procedure allows us to comparatively reveal the magnitude of the non-linear coupling between the filament and the halo and its effect on halo properties.
}
    \label{fig:sketch}
\end{figure*}

We show an example of splicing in \cref{fig:splicing_fig}, where we take a $\SI{5}{Mpc/\hred}$ sphere from randomly generated ICs $\boldsymbol{A}$ in a $\SI{50}{Mpc/\hred}$ box and splice it into another randomly generated set of ICs $\boldsymbol{B}$ to create a spliced potential field $\boldsymbol{F}$. We then plot the original ICs subtracted by the new spliced ICs ($\boldsymbol{A} - \boldsymbol{F}$) to show where the field quantities are the same. We clearly see that both set of ICs are the same within the splicing radius for density, velocity, and potential, confirming the method's effectiveness. Far away from the spliced region, the field ($\boldsymbol{A} - \boldsymbol{F}$) is Gaussian with $\sigma_{\boldsymbol{F}}^2=\sigma^2_{\boldsymbol{A}} + \sigma^2_{\boldsymbol{B}}$, as expected.
Conversely, we show in \cref{appendixg:contamination} $(\boldsymbol{B} - \boldsymbol{F})$. This highlights that the spliced field outside of the spliced region converges quickly but not immediately towards $\boldsymbol{B}$. As expected \report{from constrained field theory \citep[see e.g.][]{Bertschinger1987,Hoffman1991,vanDeWeygaert1996,Feldbrugge2023}}, the smoother the field (from density to velocity to potential), \report{the larger the power spectrum on large scales and, hence,} the longer range over which it remains correlated.
\report{A practical consequence is that, across the halos chosen in this paper, the large-scale environment may be affected by which precise halo has been spliced and where, limiting our ability to attribute which changes in the environment have primarily impacted the halo. Although unavoidable, the required change to the environment thus reflects the correlated nature of $\Lambda$CDM ICs, which our method conserves.
}

\subsection{Simulation setup}\label{subseq:Simulations}

Our numerical experiment is designed to splice (or `insert') a halo at a given location with respect to a large cosmological filament.
This entails selecting objects at low-$z$ in order to build a new set of ICs at high-$z$ for subsequent simulations.
To that end, we first run dark matter-only (DMO) simulations with uniform resolution (so called `volume' simulations) of $512^3$ particles in a box size of \SI{50}{Mpc/\hred}.
Those serve as repositories to identify possible halo and filament candidates.
We generate all initial conditions with \texttt{genetIC} \citep{geneticCode} using cosmological parameters from the Planck data release (\citet{planck2018data}; $\Omega_{\rm m}=0.3158$, $\Omega_\Lambda=0.6842$, $\sigma_8=0.8117$, $n_{\rm s}=0.9660$, $h=0.6732$) and linearly evolved to $z=75$ using the first order Zeldovich approximation \citep{ZeldovichApproximation}.
We then run the simulations down to $z=0$ with the adaptive mesh refinement code \texttt{RAMSES} \citep{ramsesCode}, where particle dynamics are computed using a multi-grid particle-mesh method and the Poisson equation is solved with a conjugate gradient solver. Grid refinements are performed when the number of particles in a cell exceeds 8 and, by $z=0$, the simulation reaches a best physical force resolution of $\sim \SI{800}{pc/\hred}$ for the volume simulations, while the subsequent zoom simulations reach $\sim \SI{400}{pc/\hred}$.

We generate outputs at time intervals of \SI{135}{Myr} starting from \SI{0.35}{Gyr}, resulting in 100 outputs. Dark matter halos are identified and cataloged using the \texttt{rockstar-galaxies} halo finder \citep{rockstarCode}, retaining all halos and subhalos with more than \SI{e3}{particles} ($\gtrsim \SI{e10}{\Msun}$). All halo quantities such as mass, angular momentum, and morphology are calculated using \texttt{rockstar-galaxies} with merger trees constructed using the \textsc{yt} \citep{yt} and \textsc{tangos} \citep{tangos,pontzenPynbodyTangosVersion2023} packages. A more comprehensive description of derived quantities is given in \cref{seq:Analysis}.

\subsection{Halo and filament selection}\label{subseq:Halo and filament selection}

For our experiment, we craft initial conditions to replace a relatively empty region near a filament with the Lagrangian patch of a halo, while systematically varying their separation.
We illustrate the process in \cref{fig:sketch}. \textbf{(A)} We first identify a halo and a filament; we detail below our selection procedure.
\textbf{(B)} We then trace back both to their respective ICs, and \textbf{(C)} roll the filament's ICs to place them near the halos' position. \textbf{(D)} We repeat the process 9 times, systematically varying the distance between the halos' Lagrangian patch and the filament's.
Hereafter, we refer to the basis vectors of the box as $\{e_x, e_y, e_z\}$, with slices in \cref{fig:sketch} done in $e_z$.
While splicing is limited to $512^3$ for numerical reasons, we obtain an effective resolution of $1024^3$ through a third-order interpolation of the density field, allowing for an effective mass resolution of $m_{\rm dm} = \SI{1.2e7}{\Msun}$.

Let us now detail our filament selection.
Filaments can be thought of as cylindrical structures, where we define the vector going through the long axis of the cylinder as $e_3$ (see \cref{fig:sketch}), typically defined as the direction of last collapse. We first identify visually prominent filaments at $z=0$ with approximate positions obtained by decomposing the box into $\SI{0.5}{Mpc}$ slices in $e_z$, looking at which slice embeds the filament.
We identify a large filament, of length on the order of $\SI{10}{Mpc}$, with its direction $e_3$ lying almost perfectly in $e_x$.
While this approach is good enough for splicing as the position of the filament may shift as a consequence of the splicing operation, a more robust analysis of the filament is performed later on using \texttt{DiSPerSE} \citep{disperseCode} when calculating filament direction and separation between halo and filament. This is detailed in \cref{subseq:Filament finding}.

Conversely, we select halo candidates from a $z=0$ snapshot. We choose halos within $1.5<M_{\rm halo}/\SI{e12}{\Msun}<3.5$, where halos masses $M$ are determined from the sum of all mass within their virial radius $r_{\rm vir}$. We select halos which are isolated, with their distance to any other halo of similar mass or higher being at least several $\si{Mpc}$. We choose isolated halos for our experiment as the effects of a changing environment are easier to quantify if their evolution is not dominated by nearby massive halos.
The halo's particles are traced back to the initial conditions forming the Lagrangian patch.
We select the region contained in the convex hull of those particles.
Indeed, the patch is, in general, not a simply connected topological object (it is analogous to Swiss cheese).
We further extend the hull by 4 pixels (equivalent to $\SI{0.4}{Mpc}$ in physical space for a $512^3$ resolution box on a $\SI{50}{Mpc/h}$ volume) to ensure we include regions that are just collapsing at $z=0$: this is what we define as the `spliced region' for this halo.

\begin{figure*}
    \centering
    \includegraphics[width=\textwidth]{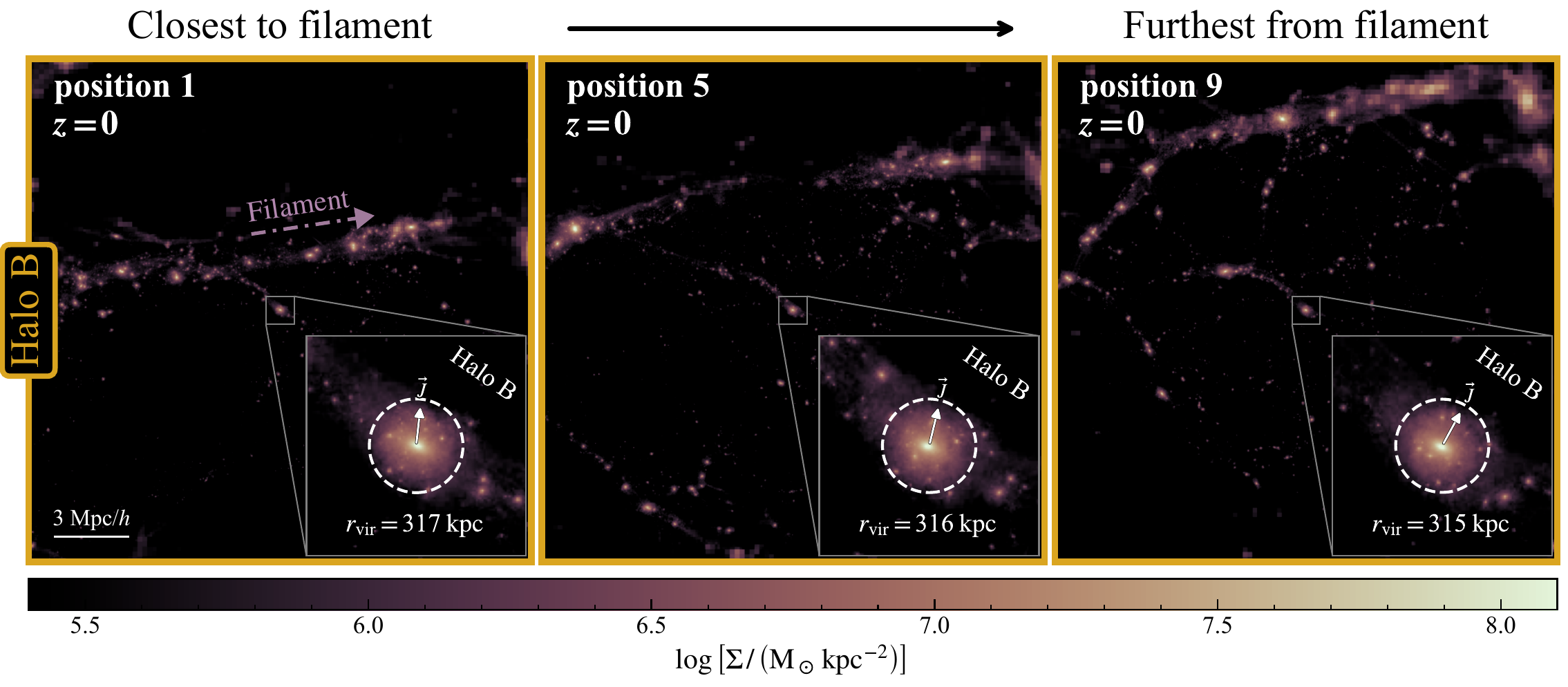}
    \caption{Density projections in $e_z$ on a \SI{5}{Mpc/\hred} thick slab centered on the halo for three positional variations of Halo B at $z=0$ which, from left to right, are the closest, middle, and furthest positions between the halo and filament. \report{In the left panel, we point out the location and direction of the position 1 filament.} We make a zoom of the halo in the bottom right corner, and plot both the virial radius $r_{\rm vir}$ as white dashed circles \report{and the angular momentum vectors $\vec{J}$ as arrows from the halo center.} The halo-filament distance is clearly increasing from left to right. Despite this drastic change of environment, the halo remain relatively intact and most of its substructure can be matched, albeit at slightly different positions. Density projections for other halos can be found in \cref{appendix:all_projections}.
    The comparison between the properties of the halo at different positions reveals the specific role played by the cosmological filament in setting them.
    }
    
    \label{fig:3_positions}
\end{figure*}
We then splice each of the selected halo at given distances from the filament.
To that end, we vary the distance in $\mathbf{e}_y$ between the halo and the filament Lagrangian patches over a large range. We keep the smallest separation in which both the halo and the filament exist at $z=0$; indeed, we don't know beforehand how close the halo can be without destroying the filament\footnote{As we splice the halo closer to the filament, the latter either disappears completely or becomes bent into an arch or other complicated morphology. We emphasize this is an expected outcome of correlated ICs.}. We then refine our method by starting from the closest position which did not destroy the filament and systematically move the halo further away at an interval of $\SI{0.75}{Mpc}$ until we reach 9 positional variations. Our full simulation suite comprises 5 halos, denoted alphabetically, where each halo has been spliced 9 times for a total of 45 halo-filament pairs.\footnote{The splicing for the last positional variation of halo E did not converge within a reasonable CPU time; we thus reject it from our sample.} The mean mass, radius, specific angular momentum, and alignment angle with the filament for each halo is given in \cref{table:values}. We plot the density projections for the closest, middle, and furthest spliced configurations of Halo B in \cref{fig:3_positions}, demonstrating the power of splicing by showing what is essentially the same halo in a varying environment. We also show a similar plot for the other 4 halos in \cref{appendix:all_projections}.

In summary, we select a filament at $z=0$ and trace it back to its Lagrangian patch in the ICs. We do the same for a MW-mass halo in another simulation, and splice in the ICs that halo's Lagrangian patch at 9 incrementally increasing distances from the proto-filament, covering a distance range of \SI{6}{Mpc}. We repeat this step for 4 other halos, which are all within a few \SI{e12}{\Msun} in mass from one another.

\begin{table}
    \caption{Mean values of virial mass, virial radius, specific angular momentum, and alignment angle across all positional variations for each halo at $z=0$. See \cref{seq:Analysis} for the definitions of those quantities.}
    \begin{tabular}{l|cccc}
    Halos & $M$ $\left[\SI{e12}{\Msun}\right]$ & $r$ $\left[\si{kpc}\right]$ & $|\mathbf{j}|$ $\left[\si{kpc.km^{-1}.s^{-1}}\right]$ & \multicolumn{1}{c}{$\cos\theta_\Jfil$} \\ \hline
    A     & 3.5                               & 400                        & 3700                                                       & 0.37                                      \\
    B     & 1.7                               & 320                        & 1900                                                       & 0.24                                      \\
    C     & 2.6                               & 360                        & 2200                                                       & 0.72                                      \\
    D     & 2.7                               & 370                        & 3500                                                       & 0.48                                      \\
    E     & 2.7                               & 370                        & 4400                                                       & 0.26                                      \\
    \end{tabular}
    \label{table:values}
\end{table}
\section{Analysis}\label{seq:Analysis}

This section describes how we take a finished spliced simulation at $z=0$ and quantify both the filament and halo through the use of \texttt{DiSPerSE}, described in \cref{subseq:Filament finding} and \texttt{rockstar-galaxies}, introduced in \cref{subseq:Simulations}.

\subsection{Filament finding}\label{subseq:Filament finding}

Selecting the filament and positioning it during splicing relied on visual inspection. In this section, we present a more principled measurement of the position and orientation of the filament to allow precise measurement of the alignment of each halo with the filament's spine. We extract the filament spine using the code \texttt{DiSPerSE} \citep{disperseCode}.
We first project the density field at $z=0$ onto a $256^3$ grid over the whole volume of $(\SI{50}{Mpc/\hred})^3$, normalized by dividing the standard deviation of the field, then smoothed by a Gaussian filter with $\sigma = \SI{1}{pixel}$. Since we are only interested in the filament for this analysis and the computational cost of \texttt{DiSPerSE} is high, we limit its input to the $[0.25, 0.75]^3$ (unitary) region which fully contains the filament. Following the approach by \citet{Espinosa2023, Espinosa2024}, we use the \{\texttt{-forceloops, -robustness, -manifolds, -upSkl, -cut}\} flags when running the \texttt{mse} command, with a cut on the field of 7. Since we work in a subbox, the field is not periodic so we set the \texttt{-periodic} flag to 0. When converting the skeleton with the \texttt{skelconv} command, we use the \{\texttt{-breakdown, -smooth}\} flags with a smoothing factor of 1. \texttt{DiSPerSE} outputs the structure of filaments as a set of connected segments.

The resulting spine tracks visually the filament. However, we noticed small variations from one spliced realization to another in the angles between different contiguous segments, as expected from slightly modified filaments due to the spliced operation and from Poisson noise in the spine construction. To ensure a robust filament direction across different realizations, we
estimate the filament's direction using a few segments of the spine, making it less subject to the given orientation of any one segment. The filament's direction is then found through a 3D line fitting of the segment positions using singular value decomposition (SVD). SVD allows us to find the eigenvectors of the matrix $A^TA$ for a set of points $A$. If $A$ is the set of vertices making up a filament, then the first eigenvector of $A^TA$ is the spine direction $\mathbf{e}_3$ along the filament. We define the distance $d_{\rm fil-halo}$ from the halo to the filament as the minimum distance between the filament's segments and the halo center.

\subsection{Halo spin-filament alignment}\label{subseq:Halo spin-filament alignment}

The angular momentum of the halo is computed in \texttt{rockstar-galaxies} from all bound particles $N$ within the virial radius $r_{\rm vir}$ as
\begin{equation}
    \mathbf{J} = \sum^N_{i=1} m_i (\mathbf{r}_i \times \mathbf{v}_i),
    \label{eq:angular_momentum}
\end{equation}
where $m_i$ is the mass of the particle, and $\mathbf{r}_i$ and $\textbf{v}_i$ are respectively the position and velocity of the particle with respect to the halo's center of mass.
We define $\theta_\Jfil$, the angle between the halo's angular momentum $J$ and filament direction $e_3$ following \citet{CosmicBalletIII}, as
\begin{equation}
    \cos{\theta}_\Jfil = \left|\frac{\mathbf{J}\cdot\mathbf{e}_3}{|\mathbf{J}|}\right|.
    \label{eq:cos_theta}
\end{equation}
This quantity in three dimensions is uniformly distributed between 0 and 1 among samples of isotropically drawn vectors, allowing to easily extract the presence of a directional bias in a sample of vectors. If halos had no correlations in angle with their nearby filaments, then we would expect the mean of the distribution to be  $0.5$.  If a population of halos tends to be preferentially parallel (perpendicular) to their nearby filament, their mean is above (below) $0.5$.

\subsection{Morphology}\label{subseq:Morphology}

Morphological parameters are computed in \texttt{rockstar-galaxies} by first computing the mass distribution tensor $M_{ij}$ for all particles $N$ within $r_{\rm vir}$
\begin{equation}
    M_{ij} = \frac{1}{N} \sum_N x_i x_j,
\end{equation}
assuming constant mass across all particles. The eigenvalues of $M_{ij}$, sorted in decreasing order, are then the square of the ellipsoidal axes $a$, $b$, and $c$ of the halo. \texttt{rockstar-galaxies} outputs the axis ratios $b/a$ and $c/a$, as well as the components of the primary ellipsoid axis vector $\mathbf{A}$. Similarly to how we calculated the alignment between angular momentum and filament direction, we define the primary ellipsoidal axis and filament direction as
\begin{equation}
    \cos{\theta}_\Afil = \left|\frac{\mathbf{A}\cdot\mathbf{e}_3}{|\mathbf{A}|}\right|.
    \label{eq:cos_A}
\end{equation}
As for the spin alignment, any departure from a mean of $0.5$ indicates a preferential shape alignment with respect to the filament.

\subsection{Other parameters}\label{subseq:Other parameters}

We briefly introduce here other interesting and important halo quantities used in our analysis of the spliced halos. Similarly to the previous quantities, these were also calculated using $\texttt{rockstar-galaxies}$.

The virial radius $r_{\rm vir}$ is calculated from the spherical overdensity where the mean density equals 200 times the background density. The mass $M$ is then the total mass of all structure and substructure within the virial radius. The virialization parameter $E_{\rm kin}/E_{\rm pot}$, the ratio between the kinetic and potential energy of the halo, is a metric for the dynamical relaxation of the halo. From the Virial theorem, a fully virialized system is one which has twice of its kinetic energy stored as potential $2E_{\rm kin} = E_{\rm pot}$. The maximal rotational velocity $v_{\rm max}$ is the maximum value of the halo's velocity profile. Total specific angular momentum $\mathbf{j}$ is the angular momentum divided by halo mass. The  spin parameter \citep{BullockSpin} $\lambda = j/(\sqrt{2}v_\mathrm{vir}r_\mathrm{vir})$ with $v_\mathrm{vir}$ being the virial velocity, measures the percentage of particles that, if placed at the virial radius, would be on circular orbits.

\subsection{Quantifying numerical noise}\label{subseq:Noise}

Quantities extracted from our simulations may be affected by different numerical noises (numerical integration, halo finding, etc.).
To determine whether variations on measured quantities of the halos across realizations are physical or merely due to numerical noise, we estimate the latter as follows.
We initialize 15 sets of identical ICs from the first positional variation of Halo A (whose $z=0$ density projection is shown in the top left panel of \cref{appendix:all_projections}), and superimpose Gaussian noise on the initial particle positions $x_i$ and velocities $v_i$ across all ICs. The new particle positions $x_{i,\rm noise}$ and velocities $v_{i,\rm noise}$ are then
\begin{align}
    x_{i,\rm noise} &= x_i + 10^{-7}\times \mathcal{N}(0, \sigma_{x_i}),\\
    v_{i,\rm noise} &= v_i + 10^{-7}\times \mathcal{N}(0, \sigma_{v_i}),
\end{align}
where $\mathcal{N}$ is a normal distribution, $\sigma$ is the standard deviation across the entire field and the factor $10^{-7}$ is such that the perturbations are close to single precision float. We then run a suite of 15 simulations from the new ICs, generate halo catalogues and calculate halo properties, and measure the standard deviation of each quantity across the 15 simulations at $z=0$.

Since the method is the same here as it is for the spliced halos, the uncertainties arising from the method should apply to both. If the scatter on quantities of the spliced halos is larger than from the 15 noise runs, then they should be as a consequence of the evolution of structures forming outside the spliced region.

\section{Results}\label{seq:Results}

In this section, we present results from our spliced halos. We particularly focus on differences in halo properties between several positional variations of the same halo, to see how the non-linear evolution of an incrementally differing environment has impacted the overall evolution of the halo.

To begin, lets revisit \cref{fig:3_positions} showing the density projections of three positional variations for halo B.
In the reference volume simulation of the halos, Halo B is embedded within a small filament, which has persisted through the splicing operation and even connects to the main filament for close variations. One unexpected but minor bi-product of splicing is apparent in the middle plot of \cref{fig:3_positions} (corresponding to the 5th or \say{middle} positional variation). The main filament has been broken up leaving a void in the center, although this is the only variation where the filament has not fully survived.

\begin{figure}
    \centering
    \includegraphics[width=0.48\textwidth]{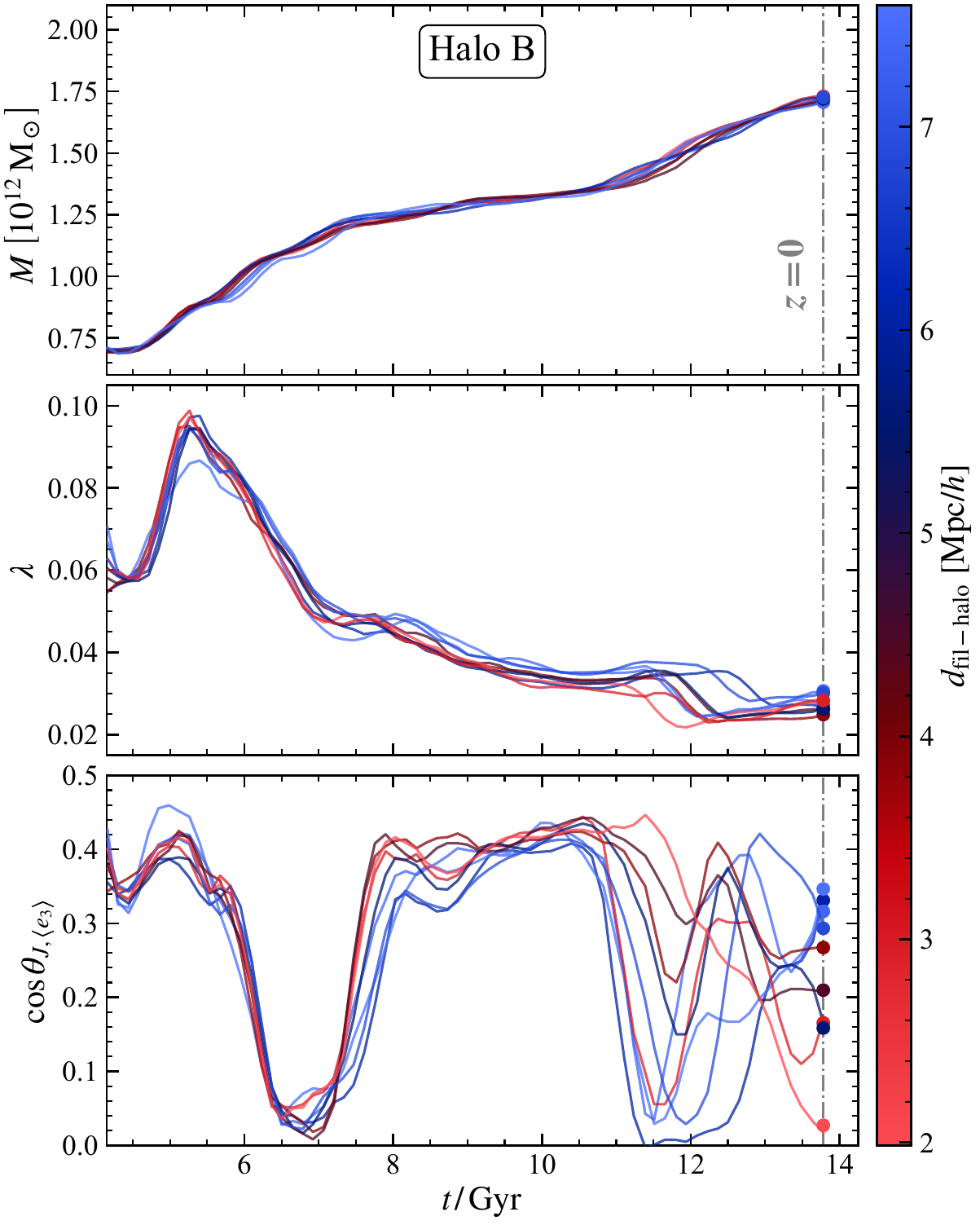}
    \caption{The time evolution of Halo B's \textbf{(top)} mass, \textbf{(middle)} spin, and \textbf{(bottom)} orientation with respect to the mean filament spine direction $\langle e_3 \rangle$, parametrized as $|\cos{\theta_{J, \langle e_3\rangle}}|$, for each spliced positional variation (9 in total) colored as their $z=0$ distance to the filament. \report{$\langle e_3 \rangle$ is the mean direction across all measured filament directions of a halo's spliced variations, so a constant vector with respect to the simulation grid. A fixed-grid vector allows us to show the similarity in orientation of the halo (with respect to the grid), instead of the early evolution being dictated by how the splicing halo has altered the filament direction.} The lines are smoothed using the Savitzky–Golay algorithm for visualization. The vertical dotted-dashed line marks present day.
    For this halo, non-linear couplings to the large-scale environment have a minor effect on final mass, a small one on spin magnitude, and a significant one on spin orientation.
    }
    \label{fig:haloB_evolution}
\end{figure}

\subsection{Time evolution of halo properties}

\begin{figure}
    \centering
    \includegraphics[width=0.48\textwidth]{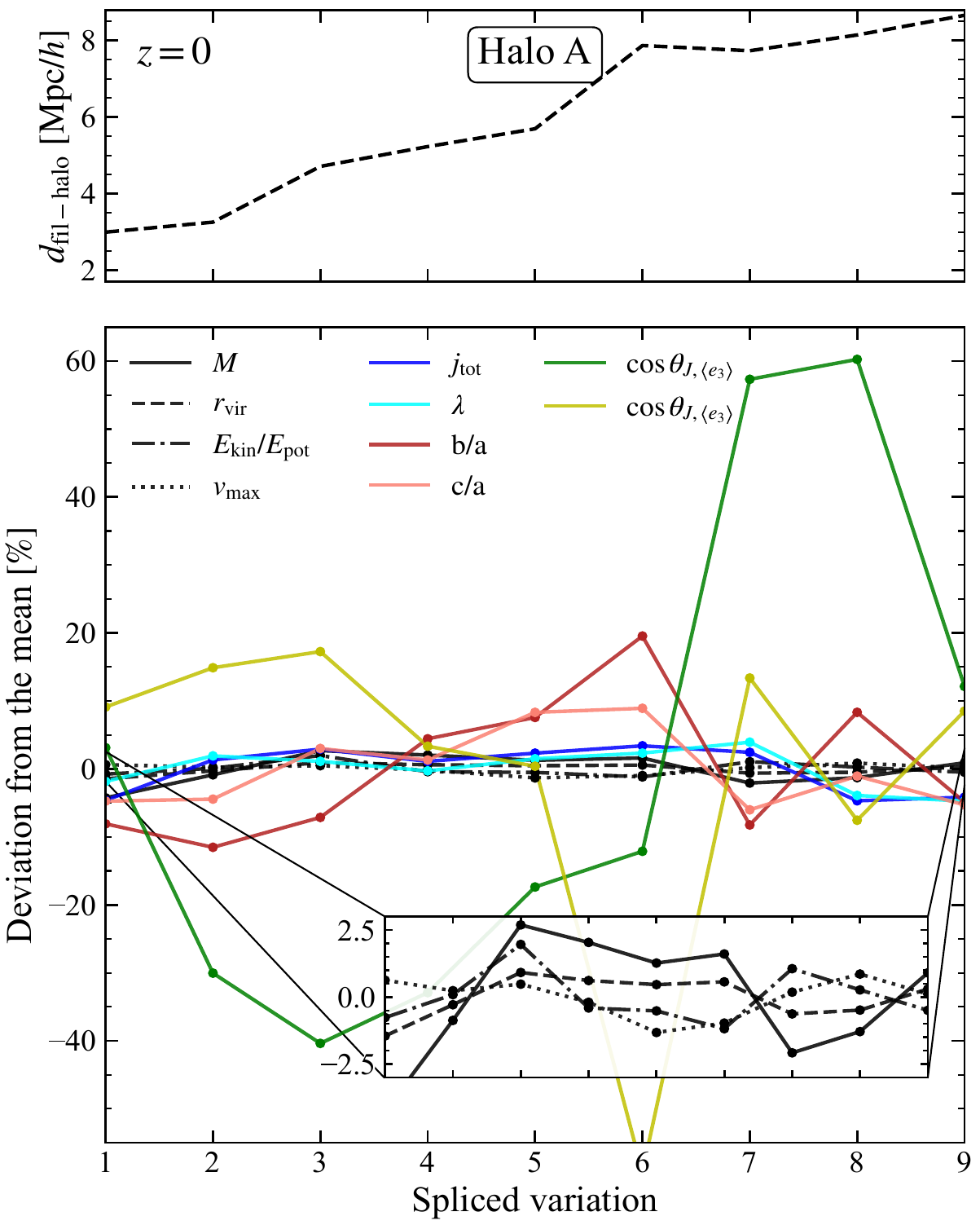}
    \caption{
    The \textbf{(top)} distance between halo and filament and \textbf{(bottom)} scatter on halo quantities, specifically mass, virial radius, virialization metric, maximal velocity, specific angular momentum, spin, morphological parameters, and orientation parameters at $z=0$ \report{with respect to the mean value,} as a function of positional variation to the major filament for Halo A. \report{Since the changes in mass, virial radius, virialization metric, and maximal velocity are all within a couple of percent, we include a zoomed in sub-plot on the lower right side.}
    Changing the initial distance between the proto-halo and proto-filament patches in the ICs lead to a corresponding, albeit not linearly related, change in their separation at $z=0$.
    At fixed initial density and tides, changing a halo's location affects its orientation to a large extent, its angular momentum and shape to a lesser one and has a limited effect on its virialization parameters, and the effect is modulated by the distance to the filament.
    }
    \label{fig:positional_variation}
\end{figure}

\begin{figure}
    \centering
    \includegraphics[width=0.48\textwidth]{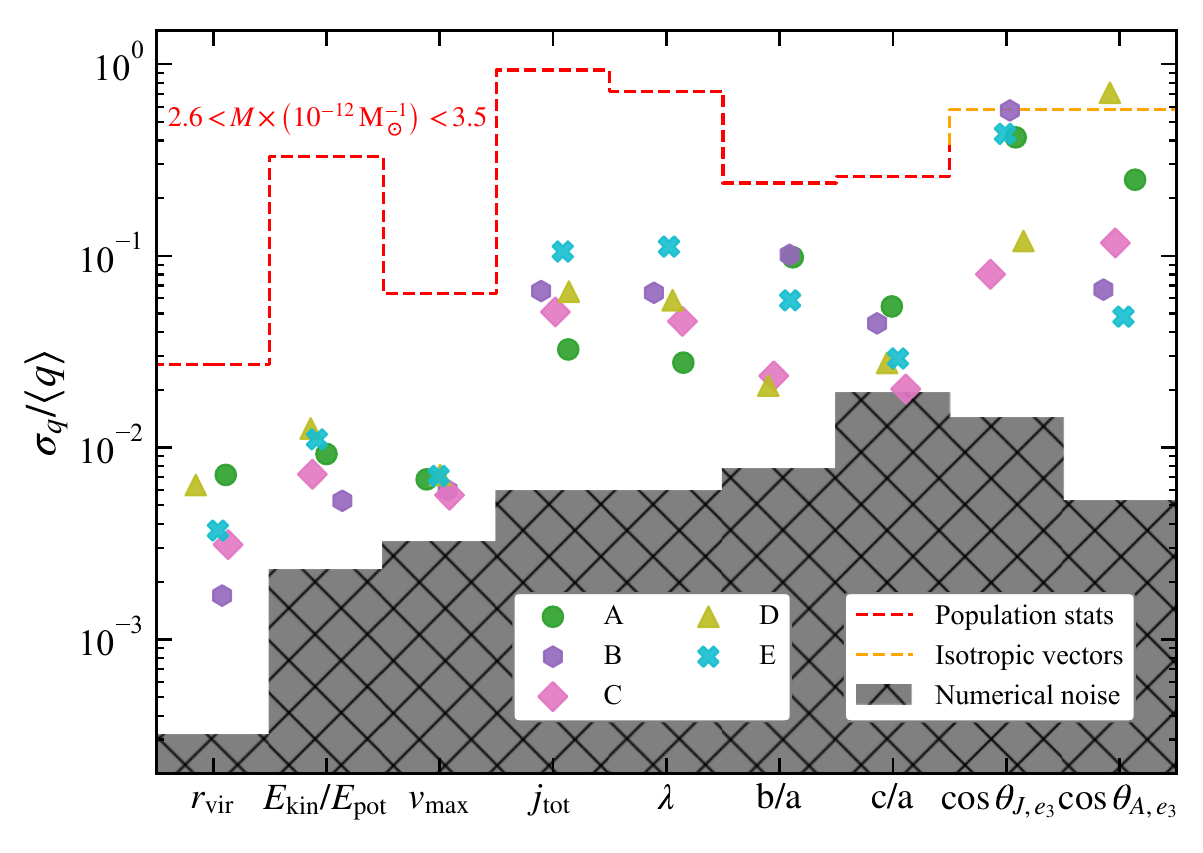}
    \caption{The scatter, parametrized as the standard deviation over the mean $\sigma_q/\langle q\rangle$, for different quantities $q$ measured at $z=0$ (from left to right, virial radius, virialization parameter, maximal velocity, total specific angular momentum, spin, morphological parameters, and orientation parameters).
    The red line shows the population-level scatter for Milky-Way mass halos ($2.6<M/(\SI{e12}{\Msun})<3.5$),
    the orange line shows the scatter for a population of isotropically oriented halos,
    the shaded gray region shows the numerical noise estimated, see the text for details.
    Colored markers represent the scatter for each halo quantity across the nine positional variations of that halo.
    The ratio between those and the numerical noise level give a signal-to-noise estimate.
    For all measured quantities (except $c/a$), the non-linear coupling to the filament leads to fluctuations several times above numerical noise levels. These fluctuations are comparable to population levels at least for orientation parameters.
    }
    \label{fig:scatter}
\end{figure}

We show the mass, spin, and orientation evolution of Halo B's 9 positional variations to the main filament in \cref{fig:haloB_evolution}. The distance between halo and filament at $z=0$ is the red to blue gradient. The difference in distance between the closest and furthest positional variation is upwards of $\SI{5}{Mpc}$. The mass, spin, and orientation evolution for other halos are shown in \cref{appendix:all_evolution}.
The mass evolution is largely the same at all epochs of mass buildup of Halo B. All other halos also follow a similar trend, with the largest deviation in mass not exceeding \SI{3}{\percent} from the mean by $z=0$. Larger deviations in mass occur during their evolutions, and can all be attributed to the timing of mergers (see e.g.\ Halo D in \cref{appendix:all_evolution}).  The persistence of mass evolution after the splicing operation shows that mass is essentially set by the initial density and tides in the Lagrangian patch, and remains largely unaffected by the non-linear evolution of halo's environments. The halo's spin parameter (middle panel) shows slightly more deviations across positional variation than in its mass, with such behavior persisting across the five different halos albeit with varying amplitude. This suggests that spin amplitude is mostly but not completely set by the initial density and tides. Finally, we show the evolution of the halo's orientation (third panel).
\report{The orientation is reported with respect to the mean orientation of the filament, $\langle e_3 \rangle$, where we take the average over the 9 positional variations.
}
While the initial evolution is the same for all positional variations of Halo B, we observe a clear decoupling at late-time ($\sim \SI{11}{Gyr}$). We find that, for Halo B, the spin of the halo becomes increasingly perpendicular to the filament as it forms closer. \report{Although the bottom plot in Fig.~\ref{fig:haloB_evolution} is with respect to mean filament direction across all spliced variations, the same trend is observed when using the filament direction of each spliced variation.}
Furthermore, we find that three out of five halos shift toward a preferentially perpendicular configuration as the halo is forming closer to the filament (see \cref{appendix:all_evolution}, bottom panels). \report{The decoupling is more pronounced for Halo A, where the spin becomes perpendicular to the filament at $\sim \SI{8}{Gyr}$.}
For all halos, the initial orientation and its evolution up to $\sim \SI{10}{Gyr}$ are very similar for all positional variations.

\subsection{The \texorpdfstring{$\boldsymbol{z=0}$}{} effects on spliced halos}

Let us now measure how positioning in the initial conditions translates into positions at $z=0$. We show in the top panel of \cref{fig:positional_variation} the relation between positional variation, with 0 being the closest configuration and 8 the furthest away, and final distance for Halo A.
The halo's final distance to the filament increases as it is spliced further from the filament in steps of \SI{0.75}{Mpc} (see \cref{appendix:all_deviation} for similar trends in the four other halos).
While the relation is not perfectly linear, as expected since the splicing operation may affect the filament's structure and position, it allows us to explore how proximity to the filament affects the halo's properties.
We then measure, for each halo, the mean over its 9 configurations of each quantity listed in \cref{seq:Analysis} at $z=0$, $\langle q\rangle$.
The bottom plot of \cref{fig:positional_variation} shows the ratio between each quantity and its mean, $q/\langle q\rangle$, as a function of positional variation as a monotonously increasing proxy for distance at $z=0$. Similarly to Halo B, we find a trend of increasing alignment of the halo's spin with the filament spine, $\cos{\theta_{J,e_3}}$, as the distance to the filament increases, although the trend is not monotonous.
As in \cref{fig:haloB_evolution}, we identify quantities that are insensitive to the distance to the filament (mass, virial radius, $E_\mathrm{kin}/E_\mathrm{pos}$ and $v_\mathrm{max}$) and some that are mildly sensitive to it ($j_\mathrm{tot}$, $\lambda$ and shape parameters). Again, shape and spin alignments show a strong evolution with distance.

Our analysis has thus far focused on individual halos, revealing that differences arise at late time and are modulated by the distance to the filament. Let us now quantify the amplitude of these variations and their significance for all five halos. To that end, we measure for each halo the standard deviation $\sigma_q$, \report{of a quantity $q$
\begin{equation}
    \sigma_q^2 = \frac{1}{N}\sum_{k=1}^N \left(q_k - \langle q \rangle\right)^2,
\end{equation}
over the $N=9$ positional variations.}
We similarly measure the standard deviation and mean value obtained from the fifteen control simulations described in \cref{subseq:Noise} as well as for a sample of halos with masses $2.6<M/\SI{e12}{\rm M_\odot}<3.5$ from a volume simulation. This allows us to estimate numerical noise level on the one hand, and the amplitude of population fluctuations on the other.
A proper measurement of the orientation parameter signal would entail analysing all filaments in a volume simulation; this goes beyond the scope of this paper. Instead, we assume an isotropic distribution for which $\sigma_{\cos \theta}=1/(2\sqrt{3})$ and $\langle \cos\theta\rangle =1/2$.
We plot the ratio $\sigma_q/\langle q\rangle$ for all five halos in \cref{fig:scatter} (colored diamonds) together with the population-level (red dashed line), an isotropic distribution (cyan dashed line), and the numerical noise (gray shaded region) .

There are three distinct groups in \cref{fig:scatter} which halo quantities fall under, depending on their absolute scatter. The first group are quantities with $\sigma_q/\langle q\rangle \lessapprox 10^{-2}$ corresponding to less than \SI{1}{\percent}  deviation from the mean. These include the virial radius, virialization, and maximal velocity. The second group are quantities of moderate variance, $10^{-2} \lessapprox \sigma_q/\langle q\rangle \lessapprox 10^{-1}$, between $1\%$ and \SI{10}{\percent} deviations. Quantities in the second group are angular momentum and morphological parameters. The final group concerns the quantities of largest scatter $10^{-1} \lessapprox \sigma_q/\langle q\rangle$ with deviations reaching up to \SI{80}{\percent}. The orientation parameters fall under the final group. We will now discuss in further details each group in the following sections.

\paragraph*{Radius, virialization, and maximal velocity.}
Across positional variation of our halos, we find negligible differences in virial radius, virialization metric, and maximal velocity compared to population-level statistics.
This suggests that their $z=0$ values are set by the local initial conditions (density, tides and velocities) in the Lagrangian patch, and are thus not sensitive to non-linear coupling to their environment.
Despite these variations being small relative to population statistics, we report that the virial radius significantly changes relative to numerical noise, having a scatter an order of magnitude larger than numerical noise. The same observation applies to virialization and maximal velocity, albeit to a lesser extent.

\paragraph*{Angular momentum and morphology.}
The specific angular momentum and spin of halos are closely linked quantities, and exhibit the same trends across all positional variations for all halos. We find that, although angular momentum can be modulated by non-linear coupling to the filament, the absolute $z=0$ scatter is minor for all halos except Halo E (see \cref{appendix:all_evolution} for the spin evolution). For this halo, we see fluctuations of up to \SI{30}{\percent} in $|\mathbf{j}|$ and $\lambda$ between positional variations (compared to \SI{100}{\percent} for the population level reported in e.g. \citealt{porciani_TestingTidaltorqueTheory_2002a,cadiouAngularMomentumEvolution2021}), with a trend of configurations closer to the filament having larger angular momentum than configurations far away.
We can conclude that the accumulation of angular momentum in halos can be heavily influenced by the non-linear coupling to the environment, although its extent depends on the specifics of the halo and filament. In other words, we can place a lower bound of  $\sim 10-\SI{30}{\percent}$ on the accuracy of any theory that aims at predicting the final magnitude of angular momentum from initial local density and tides alone\footnote{Such accuracy would however be a significant improvement over e.g.\ tidal torque theory.}.

The morphological parameters $b/a$ and $c/a$ closely follow each other, where $c/a$ is a suppressed (always closer to the mean) version of $b/a$ (see \cref{fig:positional_variation} and \cref{appendix:all_deviation}).
We limit our interpretation of the scatter $c/a$, as the observed scatter across positional variation is at most twice that of the numerical noise.
The scatter on $b/a$ is however much larger compared to the noise, and comes within a few times lower than the population scatter. In other words, a significant fraction of the population level statistics of $b/a$ is due to non-linear coupling to the cosmological environment rather than being set by the initial density, tides and velocities.

\paragraph*{Orientations with respect to the filament.}
Finally, orientation parameters are by far the most affected by the non-linear coupling to the environment.
We find the scatter in orientation, both in angular momentum and shape, across positional variation to range from \SI{10}{\percent} up to \SI{80}{\percent}. The signal is statistically strong (more than one order of magnitude larger than the numerical noise). These fluctuations levels have to be compared to the signal expected from a purely random distribution, for which we expect fluctuations of \SI{58}{\percent}.
This suggests that a large fraction of the scatter in the alignment of halos with respect to their environment is due to a non-linear coupling to this environment that cannot be accounted for in theories relying on a local analysis of the linearly-evolved initial density, tidal and velocity fields. The fact that halos exhibit trends with distance suggests that, far from being chaotic, these effects are instead modulated by the proximity to the nearest filament in a way that remains to be accounted for.

\section{Discussion and Conclusions}\label{seq:Discussion&Conclusions}

In this paper, we have constructed a numerical experiment that sets explicitly the linear part of the evolution of a halo, allowing us to reveal for the first time the magnitude of non-linear couplings between the halo and its environment.
We studied the response of 5 Milky-Way mass halos as they form progressively further from a large-scale cosmological filament, while fixing their initial density, tidal and velocity field fixed.

At fixed initial density and tides, our findings on the magnitude of non-linear coupling on halo properties are as follows:
\begin{enumerate}
    \item Mass and virialization parameters are relatively insensitive, with sub-percent fluctuations.
    \item Angular momentum and DM halo morphology are mildly sensitive to their environment, with fluctuations of up to \SI{10}{\percent}. For shape parameters, this represents half the population-level fluctuations.
    \item The orientation of the halo and of its angular momentum with respect to the nearest cosmological filament is highly sensitive to the cosmological environment, with fluctuations between \SI{10}{\percent} and \SI{80}{\percent}.
    For some halo, this is comparable to the population-level statistics.
    \item Fluctuations in the orientation are not chaotic but rather depend on the distance to the filament, although the trends with distance are not monotonic and vary from halo to halo.
\end{enumerate}

We have uncovered these results thanks to a novel extension of the splicing technique.
Compared to \cite{splicing}, we not only set the initial density but also initial tides of given halos and simulate them at varying distances from a large cosmological filament.
This allows us to uniquely disentangle what fraction of halo evolution is intrinsic to its initial density and tides from that driven by non-linear couplings to its environment. We find that, despite the Lagrangian patch of the halo being fixed in the initial conditions, some halo properties are significantly impacted by the evolution of the filament.

This work provides important pointers to improve theories aimed at predicting DM halo properties \emph{ab initio}.
Our findings that halo properties have significant fluctuations (compared to numerical noise for all, compared to population-level for some) places upper bounds on the accuracy that can be reached by models relying on the local information in the Lagrangian patch alone.
This notably includes
extended Press-Schechter theories (for mass growth predictions, \citealt{PressSchechter,Bond1991,mussoGettingShapeMinimal2023a}),
critical event theory (for merger predictions, \citealt{Hanami2001,Cadiou2020}),
machine-learning approaches based on the data in the Lagrangian patch \citep{Lucie-Smith2019,lucie-smith_InsightsOriginHalo_2022,brownUniversalModelDensity2022},
or tidal torque theory (for angular momentum predictions, \citealt{PeeblesSpin,Doroshkevich1970,White1984,Catelan1996,Crittenden2001}).
The latter one is particularly critical to the theoretical modeling of intrinsic alignment in Stage-IV surveys \citep[see][and references therein]{kiesslingGalaxyAlignmentsTheory2015a,lammanIAGuideBreakdown2024}.
In particular, our results demonstrate that predictions of the amplitude of the intrinsic alignment signal based on tidal torque theory \citep[see notably][]{Codis2015} are bound to miss a significant part of the signal \report{despite being remarkably successful at capturing it qualitatively}; effective-field theory approaches may be able to \report{improve its accuracy} \citep{Vlah2020}.

Our method improves upon the work from \citet{splicing} as we are now setting all initial local information of the halo allowing it to form in the same way early on. \report{Our work provides a complementary approach to the one in MIPs \citep{AragonCalvo2016}. In this work, the authors redrew high-$k$ modes in the initial conditions, keeping the low-$k$ modes fixed.
This allowed quick generations of large samples of (unconstrained) halos to measure the average coupling between large, fixed scales, and marginalized-over small scales.
Here, we take a reversed approach: we keep the small-scale fluctuations \emph{in real space} fixed\footnote{Working in real space is a requirement since the local density in a compact region depends on modes at all $k$.} and modify the large scale. This however comes at the expense of more limited statistical power.
} 

As \citet{splicing} observed, splicing only the density field of a halo can reduce its mass by over \SI{50}{\percent} from its original simulation. Furthermore, the scatter on sets of spliced halos in their study varied by around \SI{15}{\percent} from one another. This effect seems to be driven by a reduction in density at the boundary post-splicing, which suppresses the collapse of the halo. It could also be caused by varying tides; indeed the implementation from \citet{splicing} allows the potential and velocity fields to vary leading to a different assembly history. In our work, we find a scatter on mass across spliced halos to be on the order of \SI{1}{\percent}.
Interestingly, this allows us to estimate the `information budget' for setting halo mass, with $\sim \SI{85}{\percent}$ of the information being located in the density field within the Lagrangian patch, $\sim\SI{14}{\percent}$ in the tidal and velocity field within the patch, and the remaining $\sim\SI{1}{\percent}$ depending on non-linear couplings to the environment.

As introduced in \cref{subseq:Splicing the potential field}, splicing requires the environment field to change in order to set the density and velocity fields, which will alter the properties of the particles making up the proto-filament. Doing so inevitably alters the structure of the filament (as seen in \cref{fig:3_positions}), not only in mass and extent but in direction as well. The filament is affected not only by the field within the Lagrangian patch of the halo but also by its relative separation to it. The change in filament structure as a function of positional variation is gradual and, interestingly, the change is most prominent when the halo is furthest from the filament. While undesirable for the analysis of this paper, we cannot avoid this feedback loop of filament affecting the halo and halo affecting the filament due to the correlated nature of $\Lambda$CDM ICs.

The prospect of potential splicing may therefore not need be reserved to studying the objects formed in the spliced region. It may be equally useful to study how cosmic structures react to constraints imposed by splicing, such as filament formation given a specific tidal field configuration in its vicinity. As alluded to previously, we find that filament length and thickness are impacted by a change in the large-scale potential field, and are sometimes completely destroyed. To understand the trends of filament expansion and contraction, the impact of massive halos in the vicinity of filament or proto-filament regions needs to be assessed \citep{Espinosa2024}. A future study using splicing could in principle uncover how tidal configurations in the neighborhood of filaments precisely set filament morphology, \report{complementing that of} studies on a single realization of the cosmic web \citep[e.g.][]{Kugel2024}.

\report{As pointed out previously, Halo B is embedded within a small, local filament. This work has focused on the thick, dominant cosmic web filament, but one could also investigate the effect of the small tendrils. We also acknowledge that we are considering only one filament that was selected by visual identification, but that future studies should also explore the effect of filament diversity \citep[e.g.][]{Espinosa2020} on halo properties, for example by fixing the halo samples and repeating the experiment for filaments of different properties (e.g. short, long, thick, thin, etc.)}

\report{The nature of our experiment keeps small scales fixed in the halo region. Exploring how increasing the size of the spliced region (starting from the Lagrangian patch) affects the spliced halo evolution would be helpful in explaining the manner in which tides are precisely setting the halo properties.} \report{We could also extend the selection of halos to include a more complete sample of halo merger histories for \SI{e12}{\Msun} halos, spanning both early and late formation times.}

In a future study, we plan to introduce baryons to the simulations and form Milky-Way mass galaxies (Kuhrij et al. in prep.). Doing so will enable us to probe the impact of environment on galactic --~rather than DM halo~-- properties on an individual level and to disentangle it from the effect of halo mass growth. Probing quantities such as sSFR, quenching fractions, and morphology as the galaxy forms and evolves in a changing environment would help complement similar statistical studies of galaxies in different multi-scale environments \citep[e.g.][]{kraljic_GalaxiesFlowingOriented_2019,otherKraljic2020,songHaloMassQuenching2021, Espinosa2023}. One particular exciting line of research is to uncover whether the cosmological environment plays in role in setting the (weak) halo-galaxy alignment \citep{kimmAngularMomentumBaryons2011,genel_GalacticAngularMomentum_2015,Chisari2017,teklu_ConnectingAngularMomentum_2015,rodriguez-gomez_GalacticAngularMomentum_2022}.

\section*{Acknowledgements}

AS would like to thank Martin Rey, Katarina Kraljic, Santi Roca-F\`abrega, and Punyakoti Ganeshaiah Veena for their helpful discussions and insights. All simulations were run on the COSMOS supercomputer hosted at Lund University.
Parts of the computations to generate initial conditions and testing of the splicing technique were performed on resources provided by the National Academic Infrastructure for Supercomputing in Sweden (NAISS) at the National Supercomputer Centre (NSC), Link\"oping University (allocations NAISS 2023/6-91, and 2023/5-144).
OA and CC acknowledge support from the Knut and Alice Wallenberg Foundation, the Swedish Research Council (grant 2019-04659), and the Swedish National Space Agency (SNSA Dnr 2023-00164).

\section*{Data Availability}

The data used in this paper can be obtained on reasonable request to the corresponding author.

\bibliographystyle{mnras}
\bibliography{references}

\appendix

\section{Extra material}

\begin{figure*}
    \centering
    \includegraphics[width=1\textwidth]{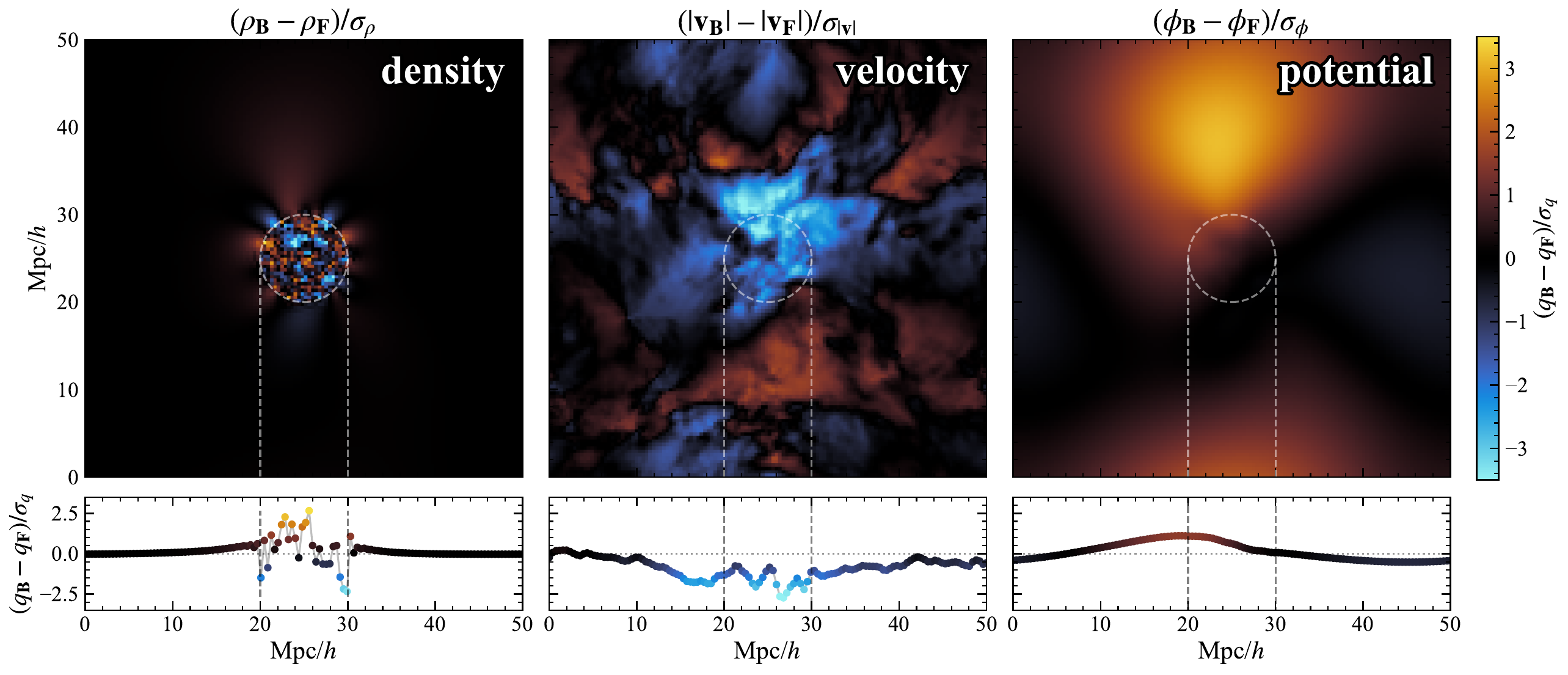}
    \caption{Difference ($\boldsymbol{B} - \boldsymbol{F}$) between two sets of cosmological initial conditions where one set $\boldsymbol{B}$ has been randomly generated, and the other $\boldsymbol{F}$ has been spliced from a different random field into the former with a sphere of radius $r=\SI{5}{Mpc/\hred}$ positioned in the center of the box. The top row shows the center slice of the box in density, velocity, and potential fields, where the colormap has been normalized by dividing out the standard deviation of each field. We take the magnitude of the velocity vectors as our scalar velocity field for visualization. The bottom row follows the horizontal line in the slice which intersects the center of the splicing region. The two dashed lines in each plot represents the spliced region. The black color indicates that the two fields are the same. The density field recombines to the original environment $\boldsymbol{B}$ on several Mpc scales, with the velocity and potential fields needing much more physical distance to recombine, on the order of the box size itself.}
    \label{appendixg:contamination}
\end{figure*}

We show how the ICs from $\boldsymbol{B}$ are affected by splicing in \cref{appendixg:contamination}, by plotting the difference ($\boldsymbol{B} - \boldsymbol{F}$). Small, local changes to the density field are shown as outflows originating at the boundary of the spliced region. Much broader changes occur in velocity and potential space.
In \cref{appendix:all_evolution}, we show the time evolution for the mass and filament-orientation of the halo for Halos A, C, D, and E.
We show the mass, spin, and orientation evolution of halos A, C, D, and E in \cref{appendix:all_evolution}.
Finally, we show the splicing-distance vs $z=0$-distance relation for halos B, C, D, and E in \cref{appendix:all_deviation}. We also plot the scatter on halo quantities as a function of splicing-distance.

\begin{figure*}
    \centering
    \includegraphics[width=0.875\textwidth]{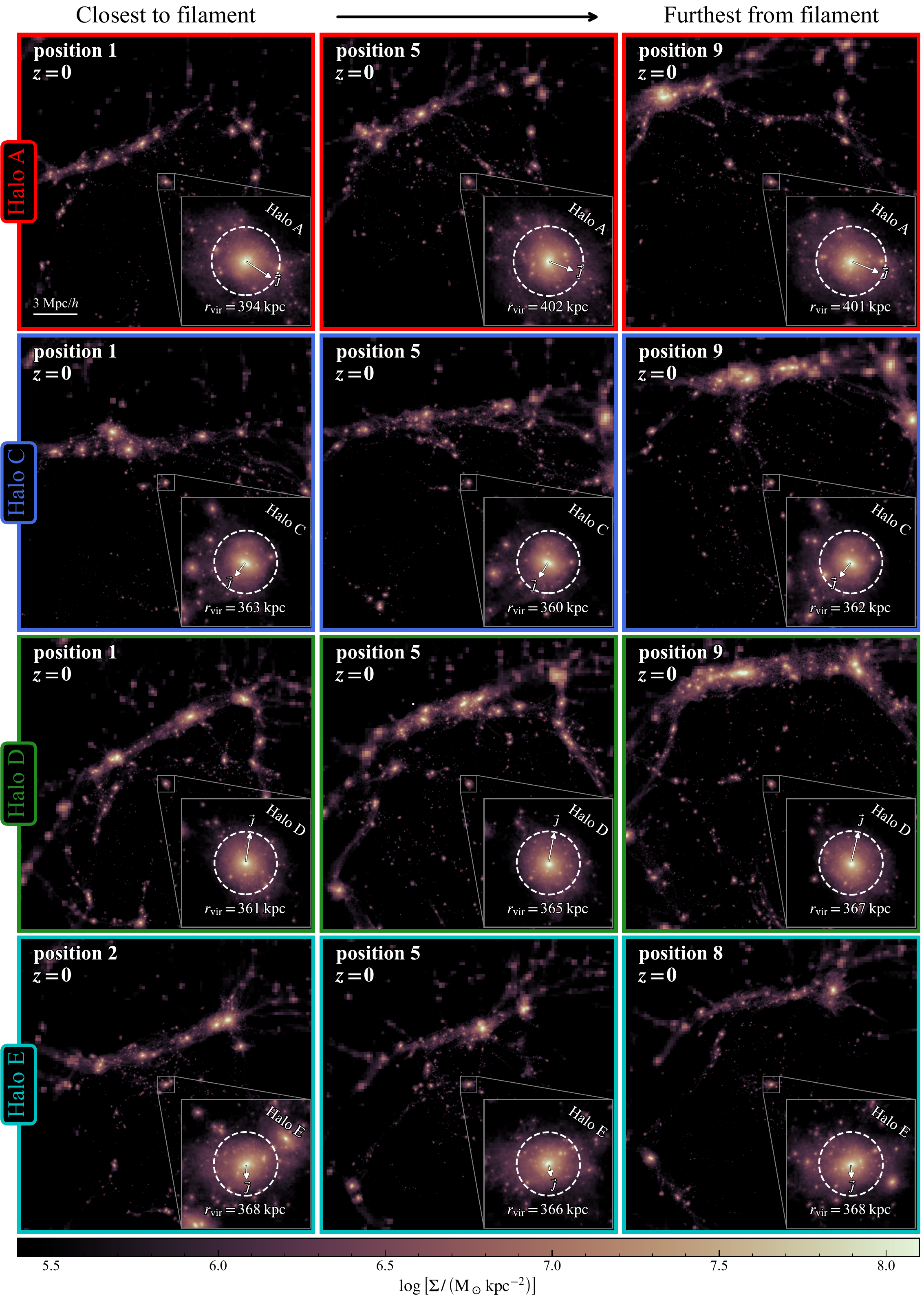}
    \caption{Density projections in $e_z$ on a \SI{5}{Mpc} slab centered on the halo for three positional variations of Halos A (red), C (blue), D (green), and E (cyan) at $z=0$ which, from left to right, are the closest, middle, and furthest positions between the halo and filament. We make a zoom of the halo in the bottom right corner, and plot both the virial radius $r_{\rm vir}$ as white dashed circles \report{and the angular momentum vectors $\vec{J}$ as arrows from the halo center.} The halo-filament distance is clearly increasing from left to right for each halo.}
    \label{appendix:all_projections}
\end{figure*}

\begin{figure*}
    \centering
    \includegraphics[width=1\textwidth]{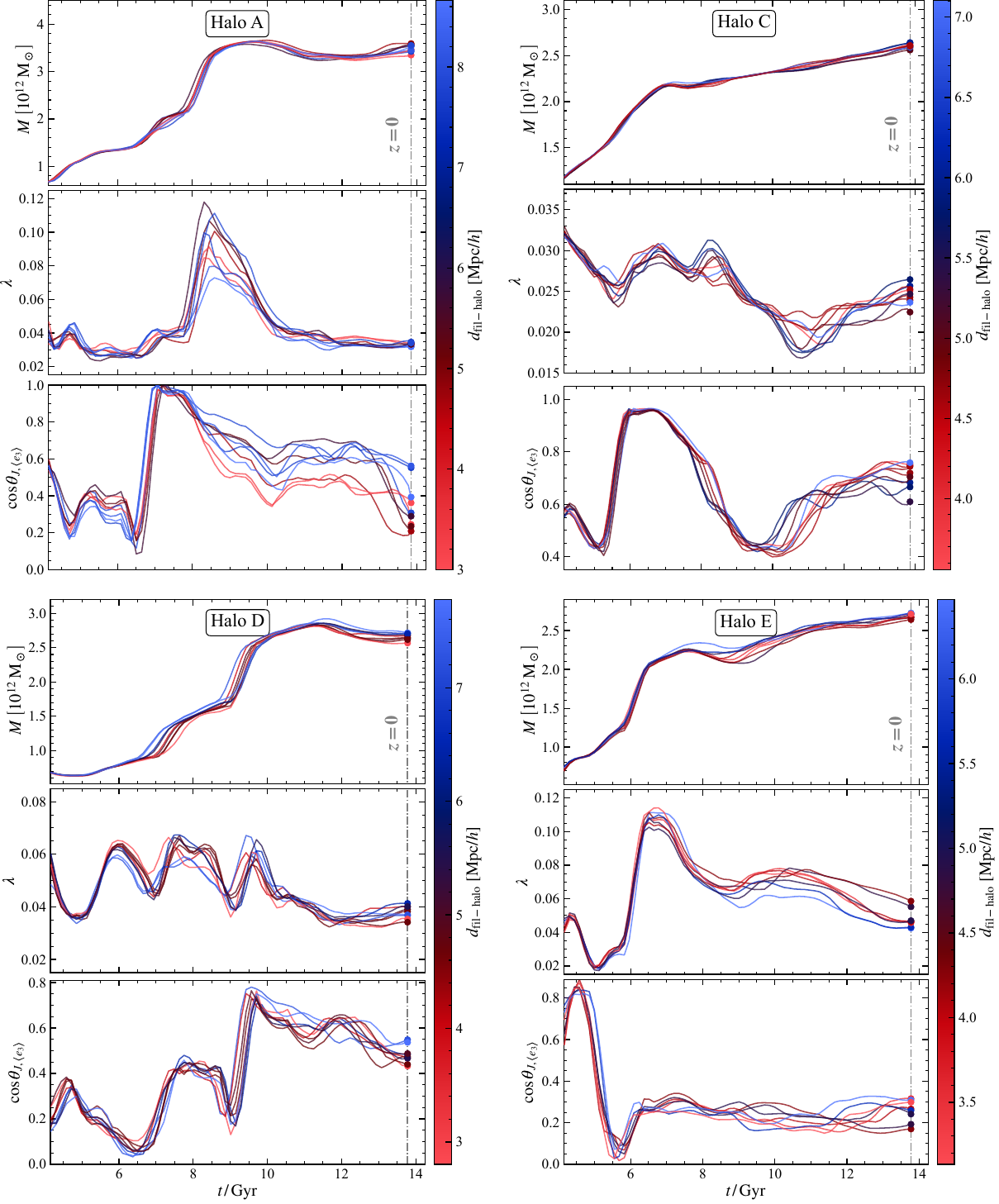}
    \caption{The time evolution of halos A, C, D, and E for their \textbf{(top)} mass, \textbf{(middle)} spin, and \textbf{(bottom)} orientation with respect to the mean filament spine direction, parametrized as $\cos{\theta_{J, e_3}}$, for each spliced positional variation (9 in total) colored as their $z=0$ distance to the filament. The lines are smoothed using the Savitzky–Golay algorithm for visualization. The vertical dotted line marks present day.}
    \label{appendix:all_evolution}
\end{figure*}

\begin{figure*}
    \centering
    \includegraphics[width=0.95\textwidth]{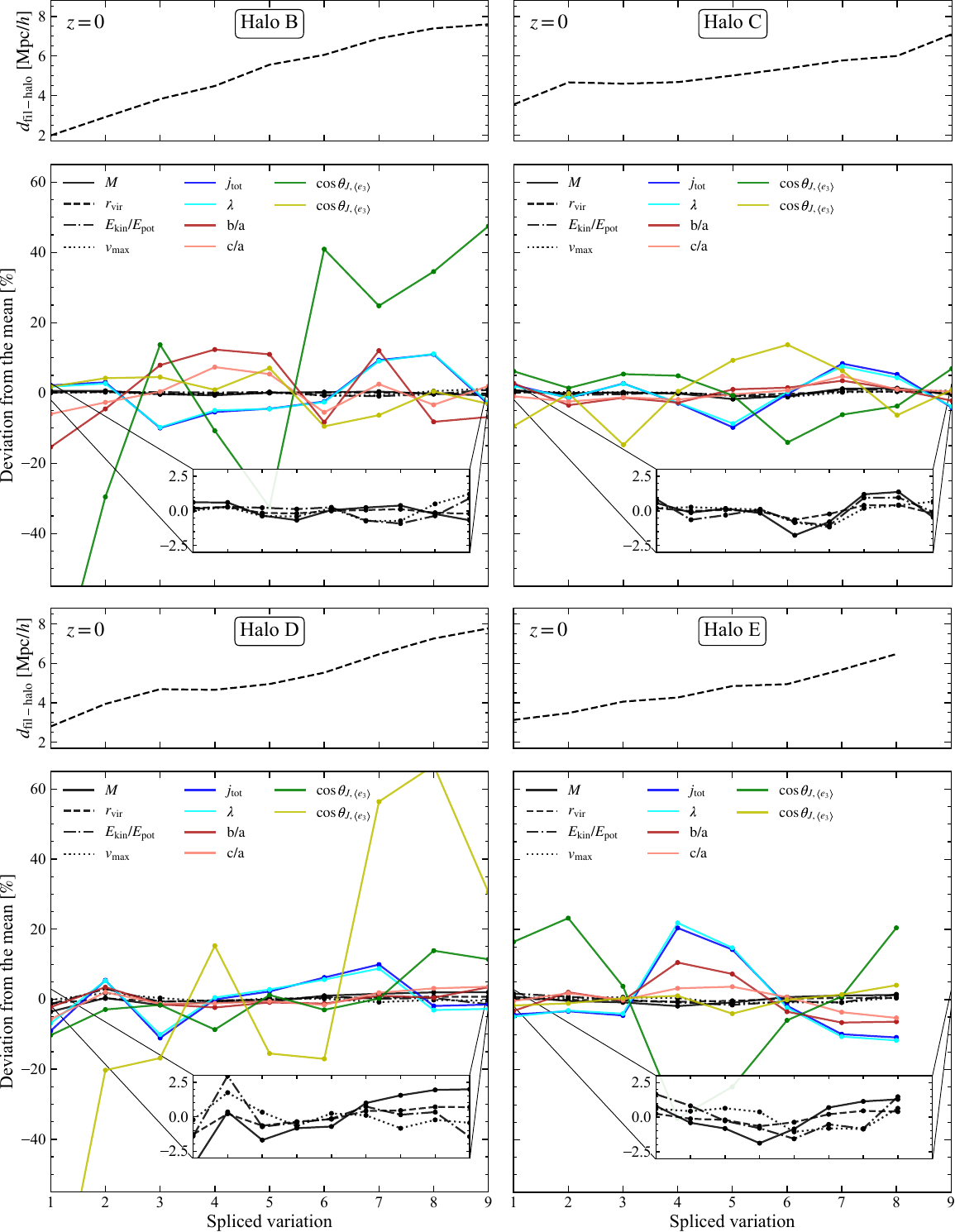}
    \caption{The \textbf{(top)} distance between halo and filament and \textbf{(bottom)} scatter on halo quantities, specifically mass, virial radius, virialization metric, specific angular momentum, spin, morphological parameters, and orientation parameters at $z=0$ as a function of positional variation to the major filament for Halos B, C, D, and E. \report{We include a zoomed in sub-plot in the lower right corners for quantities which deviate by a couple percent.}}
    \label{appendix:all_deviation}
\end{figure*}

\bsp	%
\label{lastpage}
\end{document}